\documentclass{pasj00}

\begin{document}
\SetRunningHead{Kuno et al.}{Dense Molecular Gas in Lenticular Galaxies}
\Received{2002/04/10}
\Accepted{2002/06/03}

\title{Dense Molecular Gas in Lenticular  Galaxies\thanks{This work was carried out under the common use observation program at NRO.}}

\author{Nario \textsc{Kuno}, Naomasa \textsc{Nakai}, Kazuo \textsc{Sorai}\thanks{Present 
Address: Division of Physics, Graduate School of Science, 
Hokkaido University, Sapporo 060-0810}, Kohta \textsc{Nishiyama}\thanks{
Present Address: Japan Spaceguard Center, Bisei-cho, 
Oda-gun, Okayama 714-1415} }
\affil{Nobeyama Radio Observatory\thanks{Nobeyama Radio Observatory 
(NRO) is a branch of the National Astronomical Observatory, an 
inter-university research institute operated by the Ministry of 
Education, Culture, Sports, Science and Technology.}\\ 
Minamimaki-mura, Minamisaku-gun, Nagano 384-1305}
\and 
\author{Baltasar \textsc{Vila-Vilar\'o}}
\affil{The Submillimeter Telescope Observatory\\ Steward Observatory, University of Arizona, Tucson, AZ 85721 USA}

\email{kuno@nro.nao.ac.jp}
\email{nakai@nro.nao.ac.jp}
\email{sorai@astro1.sci.hokudai.ac.jp}
\email{kota@bsgc.jsforum.or.jp}
\email{bvila@as.arizona.edu}

\maketitle

\begin{abstract}
We made CO and HCN simultaneous observations of lenticular galaxies, NGC 404, NGC 3593 and NGC 4293, and detected HCN emission in NGC 3593 and NGC 4293 as well as CO in all the galaxies. 
The $I_{\rm HCN}$/$I_{\rm CO}$ ratios were $0.025 \pm 0.006$ and $0.066 \pm 0.005$ in NGC 3593 and NGC4293, respectively, which are comparable to the late-type spiral galaxies. 
The average of the $I_{\rm HCN}$/$I_{\rm CO}$ ratios at the center of 12 nearby spiral galaxies including late-type was $0.055 \pm 0.028$.
The line profiles of CO and HCN emission showed different shape in both galaxies.
The HCN peaks were not at the systemic velocity of these galaxies, while the CO peaks were near the systemic velocity. 
These results suggest that the fraction of the dense molecular gas is high around the center in these galaxies.

\end{abstract}

\section{Introduction}

Since it has been known that even early-type galaxies contain an interstellar medium (ISM), the properties of the ISM have been examined. 
These investigations have revealed the similarities and dissimilarities of the properties of the ISM in early and late-type spiral galaxies. 
For example, lenticular galaxies contain almost an order of magnitude less molecular gas than late-type spiral galaxies (Wiklind, Henkel 1989; Thronson et al. 1989), and the molecular gas concentrates in the central region in early-type spirals (Taniguchi et al. 1994; Young et al. 1995). 
The molecular to atomic gas mass ratio is larger in early-types than late-types (Young, Knezek 1989; Sage 1993; Casoli et al. 1998). 
On the other hand, it has been shown that star formation efficiency (SFE) derived from star formation rate (SFR) and molecular gas mass are similar in early and late-type spirals (Wiklind, Henkel 1989; Thronson et al. 1989; Rownd, Young 1999).

About the central activity, it has been suggested that there exists a difference in the star formation activity and AGN with Hubble type. Ho et al. (1997b) show that H$\alpha$ luminosity of nuclei is significantly enhanced in early-type galaxies from their spectroscopic survey. 
Alonso-Herrero and Knapen (2001) also found the same tendency from the analysis of archival HST/NICMOS H-band and Pa$\alpha$ data. 
They suggest that bars affect the star-forming activity, especially in early-type spirals. 
Ho et al. (1997a) show that AGNs are found predominantly in luminous, early-type galaxies, while H\,{\footnotesize II} nuclei prefer less luminous, late-type systems. 
Furthermore, Lei et al. (2000) indicate from their analysis of a magnitude-limited sample of LINERs that the intensity of AGN activity increases with decreasing star forming contribution from late-type spiral galaxies to early-type ones. 
They suggest an evolutionary connection between AGNs and starbursts in LINERs.

Dense molecular gas is thought to play an important role for star formation. 
Moreover, concentration of dense molecular gas has been found in some of the Seyfert nuclei (e.g., Kohno 1998). 
Therefore, it is very interesting to investigate from observations of tracers of dense molecular gas whether there is the difference of density of molecular gas in the central region between early and late-type galaxies. 
The number of early-type galaxies in which tracers of dense molecular gas have been detected is still very small (Henkel, Wiklind 1997; Kohno et al. 2001). 
Since the molecular gas in early-type galaxies is expected to concentrate in the central region, it is preferred to observe nearby galaxies with a small beam size to make the beam-filling factor large. 
Therefore, we made CO and HCN simultaneous observations of nearby early-type galaxies using the 45-m telescope at the Nobeyama Radio Observatory (NRO). 
Because of the difference of the critical density for collisional excitation of CO (a few 10$^{2}$ cm$^{-3}$) and HCN ($> 10^{4}$ cm$^{-3}$) lines, the ratio of the intensity of these lines gives a rough estimation of density of the molecular gas.
We selected three lenticular galaxies (NGC 404, NGC3593 and NGC4293) in which strong CO emission have been detected (Wiklind, Henkel 1989). 

\section{Observations}

The observations were made in 1998 February and 1999 April with the 45-m telescope at NRO. 
We observed the center positions of the sample galaxies listed in table 1. 
CO($J$ = 1 -- 0) (rest frequency 115.271204 GHz) and HCN($J$ = 1 -- 0) (88.631602 GHz) lines were observed with two SIS receivers (S80 and S100) that can observe two orthogonal polarizations simultaneously. 
The pointing discrepancy between these receivers is less than 2$''$. 
The receiver backends were 2048 channel wide-band acousto-optical spectrometers (AOSs). 
The frequency resolution and total bandwidth were 250 kHz and 250 MHz, respectively, which correspond to 0.9 km s$^{-1}$ and 850 km s$^{-1}$ at 89 GHz and 0.65 km s$^{-1}$ and 650 km s$^{-1}$ at 115GHz. 
The intensity calibration was performed by chopper-wheel method. 
The system noise temperatures (SSB) including the antenna ohmicloss and the atmospheric effect were 230 K -- 270 K at 89 GHz and 600 K -- 750 K at 115 GHz. 
The telescope pointing was checked every hour by observing SiO maser emission of late type stars R-Vir, W-And and R-Leo at 43 GHz and radio continuum source 3C273. 
In the case that the pointing offset before and after an observation was larger than 5$''$, we did not use the data. 
We used a position-switching mode with the integration time of 20 sec for on- and off-source and checked the spectra of all scans.
In the case that either of CO and HCN was a bad scan, we did not use both spectra of CO and HCN of the scan.
This is because simultaneous observations are very important to compare the two lines.
Since for position switching some level of baseline offset remains, we applied linear baseline correction.

We used the main beam efficiency $\eta_{\rm mb}$ measured at 110 GHz for CO and that at 86 GHz for HCN to convert an antenna temperature $T_{\rm A}^{*}$ into the main beam brightness temperature $T_{\rm MB}$ ($T_{\rm A}^{*}$/$\eta_{\rm mb}$). 
The main beam efficiency at 110 GHz (S100) in 1998 February and 1999 April were 0.46 and 0.48, respectively. 
The main beam efficiency at 86 GHz (S80) was 0.5 in 1999 April. 
Although there are no measurements of the main beam efficiency at 86 GHz in 1998 February, we assumed that the efficiency at 86GHz was same in 1998 February and 1999 April, since the efficiency at 110 GHz measured with the receiver S80 was same in 1998 and 1999. 
The half-power beam widths measured at 110 GHz and 86 GHz were about 15$''$ and 19$''$, respectively.

\section{Results and Discussions}
\subsection{NGC 404}
NGC 404 is an S0 galaxy with central star-forming activity and classified as a LINER (Ho et al. 1997c). 
We adopted the center position determined by the optical data (Palumbo et al. 1988). 
The position is 1$''$ south from the compact UV source in the nucleus found by Maoz et al. (1995) and a few arc second east from the center of the CO map of Wiklind and Henkel (1990) obtained with Onsala 20-m telescope. 
Wiklind and Henkel (1990) found that the CO emission is not centrally peaked. 
The CO distribution shows an arc-like feature that corresponds to the dust lane seen in the optical photograph, and whose size in the CO map is about 40$''$.
Our 19$''$ beam for HCN corresponds to 220 pc and does not cover the whole arc-like structure seen in CO emission, while the bright UV sources observed by Maoz et al. (1995) including the components from young stars near the nucleus are fully covered. 
Although Wiklind and Henkel (1990) assert that the distance of the galaxy is 10 Mpc, we adopted 2.4 Mpc derived from the systemic velocity from Tully (1988). 

Figure 1a shows a CO spectrum measured by our observations. Its line shape is similar to the profile at about 10$''$ east from the center in the map of Wiklind and Henkel (1990) rather than that in the center. 
This may be due to the pointing errors of their and our measurements.
Since our adopted center position is a few arc second east from their center postion and the pointing errors are estimated about 5$''$ and 4$''$ for ours and theirs, respectively, our result does not conflict with theirs.
The column density of H$_{2}$ derived using the conversion factor of $2.3\times 10^{20}$ cm$^{-2}$ [K km s$^{-1}$]$^{-1}$ (Strong et al. 1988) is 47 $M_{\solar}$ pc$^{-2}$. 
We did not detect HCN emission from this position. 
The 2-sigma upper limit of $I_{\rm HCN}$/$I_{\rm CO}$ ratio ($I \equiv \int T_{\rm MB} dV$) is 0.042, which is still comparable with late-type spirals (Sorai et al. 2002; Kohno 1998). 
Since we observed only one position, we can not deny that we missed a clump of dense molecular gas. 
The CO peak found by Wiklind and Henkel (1990) is shifted by ($+10''$, $-10''$) from our observed position.
At least, it is necessary to observe the CO peak, since some galaxies show a HCN peak that corresponds to CO peaks out of the nuclei (e.g., NGC3593, see section 3.2).

\subsection{NGC 3593}
NGC 3593 is an S0/a galaxy that has a counterrotating stellar disk (Bertola et al. 1996). 
Hunter et al. (1989) showed that this galaxy is under going a burst of star formation and that the star-forming regions are concentrated in the small central region. 
Corsini et al. (1998) found that the star-forming regions consist of a filamentary pattern with a circumnuclear ring. 
The high resolution CO maps of the central region of NGC 3593 show three peaks (Sakamoto et al. 2000; Garc\'ia-Burilo et al. 2000). 
The central peak is located at the galactic center and other two peaks seem to be associated with the circumnuclear star-forming ring. 
Our 19$''$ beam for HCN, which corresponds to 510 pc assuming that the distance of NGC 3593 is 5.5 Mpc (Tully 1988), covers most of the H$\alpha$ ring whose diameter is about 17$''$. 
We adopted the radio continuum peak (Condon 1987) as the center and detected both CO and HCN emissions (figure 1b). 
The CO profile and intensity are consistent with that obtained with the IRAM 30-m telescope (Wiklind and Henkel 1992). The nearest point to our observed position is (10$''$, 0$''$) in the IRAM map. 
The column density of H$_{2}$ is estimated to be 199 $M_{\solar}$ pc$^{-2}$ using a conversion factor of $2.3\times 10^{20}$ cm$^{-2}$ [K km s$^{-1}$]$^{-1}$. 
On the other hand, our HCN spectrum shows a different profile from that detected by Henkel and Wiklind (1997) with the IRAM 30-m telescope. 
Our spectrum is narrower ($\Delta V$ = 100 km s$^{-1}$) and stronger ($T_{\rm peak}$ = 20 mK) than theirs ($\Delta V > 200$ km s$^{-1}$; $T_{\rm peak} <$ 10 mK). 
The difference of the beam size and the observed position may be the cause of the difference of the profiles. 
Unfortunately, because their observed position was not stated, we can not check the difference between our observed position and theirs.

It is interesting that the CO and HCN spectra show different shape. 
Since we observed these lines simultaneously, the difference is real and not due to the pointing error. 
The HCN emission was detected in the velocity range only 500 - 600 km s$^{-1}$, while the CO emission was detected in the velocity range 500 - 800 km s$^{-1}$ at the zero intensity level. 
The difference is apparent also in the FWHM.
The FWHM of the HCN profile is about 50 km s$^{-1}$, while that of the CO profile is about 130 km s$^{-1}$.
The peak velocity of the CO profile is 605 km s$^{-1}$ which is near the systemic velocity of the galaxy. 
On the other hand, the HCN spectrum does not have a peak at the velocity. 
The peak velocity of the HCN profile is 540 km s$^{-1}$.
The $I_{\rm HCN}$/$I_{\rm CO}$ ratio is 0.025 $\pm$ 0.006. The ratio of $T_{\rm MB}$ in the velocity range where HCN was detected is higher ($\sim 0.07$).

It is apparent from the velocity field in Garc\'ia-Burilo et al. (2000) that we detected HCN emission from the western peak only. 
This result means that this part has a higher fraction of dense molecular gas than in the other region within our 19" beam. 
This region coincides with the site where the CO(2 -- 1)/CO(1 -- 0) ratio is high and the most vigorous star-forming region (Wiklind, Henkel 1992; Corsini et al. 1998). 
It seems to be reasonable that the star-forming activity is high there because of the high fraction of the dense molecular gas.

\subsection{NGC 4293}
NGC 4293 is an S0/a galaxy and classified as LINER (Ho et al. 1997c). 
There is a dust lane near the center (B\"oker et al. 1999). 
H$\alpha$ emission concentrates on the nucleus except for another peak about 40$''$ east from the center (Koopmann et al. 2001). 
Our 19$''$ beam for HCN, which corresponds to 1600 pc at the adopted distance of 17 Mpc, covers the dust lane and the central H$\alpha$ peak.

We detected CO and HCN emission toward the center of NGC 4293 (figure 1c). 
The $I_{\rm HCN}$/$I_{\rm CO}$ ratio is 0.066 $\pm$ 0.005 and the ratio of $T_{\rm MB}$ at the peak of HCN is higher ($\sim 0.1$), which are highest among three galaxies we observed. 
The CO profile is quite similar to the CO(2 -- 1) spectra in Wiklind and Henkel (1989). 
The column density derived from the $I_{\rm CO}$ is 145 $M_{\solar}$ pc$^{-2}$ assuming the conversion factor of $2.3\times 10^{20}$ cm$^{-2}$ [K km s$^{-1}$]$^{-1}$. 
As seen in NGC 3593, the CO and HCN spectra show a difference in the shape. 
The HCN emission is weak at the systemic velocity and has a double peak, while the CO spectrum has a single peak near the systemic velocity of this galaxy. 
Simple interpretation of this result is that dense molecular gas traced by HCN distributes around the center and has a ring-like structure whose size is less than our beam size, while less dense molecular gas traced by CO concentrates in the center.
The dense gas is expected to associate with the star-forming activity in the center.
Usui et al. (2001) found a correlation between turn-over radius of the rotation curve from rigid rotation to flat rotation and the radius of the active star-forming region in the central regions of early-type spirals.
Since the rotation curve of this galaxy rises steeply near the center (Rubin et al. 1999), the diameter of the ring-like structure, if it exists and associates with the star-forming activity in the central region, must be small. 
The velocity separation of the double peak of HCN is about 100 km s$^{-1}$ in radial velocity. 
From the rotation curve, the velocity separation corresponds to a radius of a few arc second, or of about 200 pc. 
Therefore, if the dense molecular gas exists in the turn-over radius, the radius is expected to be about 200 pc.
These features resemble the structures seen in NGC 3593, although it seems that the distribution of the dense gas in the ring of NGC 3593 is not symmetric with respect to the center. 
It is highly desirable to get high-resolution images of CO and HCN to confirm the structures. 
Although the large inclination angle of this galaxy is a disadvantage, the velocity information will give us some clues about the spatial structure. 
Furthermore, high-resolution image of tracers of star forming regions are required to examine the relation between dense molecular gas and star formation there. 

\subsection{Comparison with later type galaxies}
As compared with the previous survey of HCN including late-type galaxies, the $I_{\rm HCN}$/$I_{\rm CO}$ ratio we obtained (table 2) is comparable to the late-type galaxies. 
Figure 2 shows the correlation between $I_{\rm CO}$ and $I_{\rm HCN}$, including the data toward the center of other galaxies measured with the NRO 45-m telescope from the literature for the comparison (Sorai et al. 2002; Kohno 1998; Kohno et al. 2002). 
The range and average of the distance of the galaxies are 1.8 - 24.1 Mpc and 8 Mpc, respectively, which are comparable with our sample (2.4 - 17 Mpc, 8.3 Mpc).
We used the data of the same telescope, since the beam size and the intensity calibration method are the same. 

The lenticular galaxies are distributed in lower side of $I_{\rm CO}$ in figure 2. 
It means that the column density of molecular gas averaged by our 19$''$ beam is lower in the lenticular galaxies. 
Since the galaxies plotted here are CO bright galaxies in each morphological type, this result indicates that the upper side of the dispersion of the column density of molecular gas at the central region is higher in later type galaxies.
The rate of concentration of molecular gas in the center may be one of the reasons of this trend.
Taniguchi et al. (1994) showed that three S0 galaxies out of four have significantly small scale lengths of radial distribution of the molecular gas.
The shorter scale length of the column density of molecular gas in early-type galaxies will make the averaged surface density within our 19$''$ beam lower, even if there is no difference in the surface density within much smaller area.

The average of the $I_{\rm HCN}$/$I_{\rm CO}$ ratio of all the data plotted in figure 2, except for the upper limits, is 0.055 $\pm$ 0.028 (1$\sigma$). 
On the other hand, the best fit with power-law is $I_{\rm HCN} \propto I_{\rm CO}^{0.9}$. 
Since the sample number is still small, the result is not conclusive. 
Especially, when we take into account the upper limit in the lower side, the power seems to be larger than 1.
It must be important to check the relation to examine the mechanism of star formation.
The star formation scenario based on the large-scale gravitational instability seems to do well to explain the relation between SFR and the surface density of interstellar gas (Kennicutt 1989). 
Most of the recent results of the observational studies about the relation between SFR and molecular gas show a non-linear relation and a power-law slope of 1.2 -- 1.4 (Kennicutt 1998; Rownd, Young 1999). 
However, the mechanism that makes stars from the molecular gas in large molecular complexes formed by gravitational instability is still an open question. 
Because star formation should begin from creation of dense molecular gas that can be traced by HCN, we may be able to examine different phases in the process of star formation, namely, from diffuse molecular gas to dense molecular gas and from dense molecular gas to stars, from the relations between $I_{\rm CO}$ and $I_{\rm HCN}$ and between $I_{\rm HCN}$ and SFR, comparing with the relation between $I_{\rm CO}$ and SFR.
 
The different shape of HCN and CO spectra in the early-types in figure 1 is very interesting. 
For late-type galaxies, the difference is not prominent (e.g., Sorai et al. 2002). 
Although the beam size of HCN is slightly larger than that of CO, it must be difficult to attribute the difference of the profiles to the difference of the beam size.
If the difference attributed to the beam size, the HCN emission must exit just out of the CO beam.
Our results imply that in NGC 3593 and NGC 4293 the fraction of dense molecular gas is higher around the center and the dense gas may be associated with a ring-like structure. 
Pogge and Eskridge (1993) found that a ring of H\,{\footnotesize II} regions is fairly common feature in S0 galaxies. 
The dense molecular gas must be associated with the star-forming activities in these galaxies as seen in NGC 3593. 
As mentioned in the previous section, the location of the star-forming region seems to be correlated with the turn-over radius of the rotation curve.
Actually, the radius of the ring-like structure of CO and H$\alpha$ in NGC 3593 corresponds to the turn-over radius of the rotation curve (Corsini et al. 1998; Garc\'ia-Burillo et al. 2000).
This relation is explained by viscosity of molecular clouds (Icke 1978; Fukunaga 1983).
Namely, in the region of the differential rotation, the molecular clouds lose angular momentum due to shear motions, while shear motions do not occur in the region of the rigid rotation.
As a result, the molecular gas accumulates in the turn-over radius.
The same relation has been observationally found in late-type galaxies (Fukunaga 1984; Nishiyama et al. 2001).
If the molecular gas is accumulated by this mechanism, it is expected that the molecular gas concentrates in a narrower region in early-type galaxies than late-type ones, since rotation curve arises more steeply in early-type galaxies than late-type.
Therefore, we speculate that the density of the molecular gas might become extremely high in such a region in early-type galaxies and active star formation might occur there, although, of course, it must depend on the total amount of the accumulated gas.
Since the sample galaxies we selected are CO bright galaxies, observations of galaxies in wider range of CO brightness are required to avoid a selection effect.
Further observations with an interferometer are also required to know the distribution of molecular gas, especially dense molecular gas, and relations with star-forming activity and dynamics of molecular gas. 

\section{Summary}
We made HCN and CO simultaneous observations of three lenticular galaxies, NGC 404, NGC 3593 and NGC 4293, using the Nobeyama 45-m telescope. 
We detected CO emission from all galaxies and HCN emission from NGC 3593 and NGC 4293. 
The $I_{\rm HCN}$/$I_{\rm CO}$ ratios are $0.025 \pm 0.006$ and $0.066 \pm 0.005$ in NGC 3593 and NGC 4293, respectively, which are comparable to that in late-type galaxies. 
The average of the $I_{\rm HCN}$/$I_{\rm CO}$ ratios at the center of 12 nearby spiral galaxies including late-type was $0.055 \pm 0.028$.
The profiles of HCN emission in both galaxies show different shape from those of CO emission. 
Namely, the HCN emission was detected from only a part of probably the ring-like structure in NGC 3593 and the HCN profile shows a double peak in NGC 4293, while the CO spectra in both galaxies have a peak at their systemic velocity. 
These results indicate that the fraction of dense molecular gas is high around the center and the dense gas associates with the ring-like structure and the star-forming activities in these galaxies.

\vspace{1pc}\par
This research has made use of the NASA/IPAC Extragalactic Database 
(NED) which is operated by the Jet Propulsion Laboratory, 
California Institute of Technology, under contract with the National 
Aeronautics and Apace Administration.


\clearpage

\begin{table}
\caption{Properties of Sample Galaxies}\label{}
\begin{center}
\begin{tabular}{ccccccc}
\hline\hline
Galaxy & R.A.(2000) & Dec.(2000) & Type$^{a}$ &
$D$ (Mpc)$^{a}$ & $V_{\rm sys}$ (km s$^{-1}$)$^{b}$ & Inclination angle (deg)$^{a}$\\
\hline
NGC 404 & \timeform{1h9m26.937s} & \timeform{35D43'3.81''} $^{c}$
& S0 & 2.4 & $-48$ & 0\\
NGC 3593 & \timeform{11h14m36.86s} & \timeform{12D49'5.57''} $^{d}$
& S0/a & 5.5 & $628$ & 69\\
NGC 4293 & \timeform{12h21m12.91s} & \timeform{18D22'57.7''} $^{e}$
& S0/a & 17 & $948$ & 66\\
\hline
\end{tabular}
\begin{tabular}{l}
{\footnotesize a: \citet{Tu1988}}\\
{\footnotesize b: \citet{de1991}. Heliocentric velocity}\\
{\footnotesize c: \citet{Pa1988}}\\
{\footnotesize d: \citet{Co1987}}\\
{\footnotesize e: \citet{Dr1976}}\\
\end{tabular}
\end{center}
\end{table}

\begin{table}
\caption{Measured intensity of CO and HCN emission}\label{}
\begin{center}
\begin{tabular}{cccc}
\hline\hline
Galaxy & $I_{\rm CO}$ (K km s$^{-1}$) & $I_{\rm HCN}$ (K km s$^{-1}$) & $I_{\rm HCN}$/$I_{\rm CO}$\\
\hline
NGC 404 & $12.7 \pm 0.5$ & $< 0.53^{*}$ & $< 0.042$\\
NGC 3593 & $54.2 \pm 1.2$ & $1.34 \pm 0.36$ & $0.025 \pm 0.007$\\
NGC 4293 & $39.4 \pm 0.8$ & $2.62 \pm 0.25$ & $0.066 \pm 0.006$\\
\hline 
\end{tabular}
\\
\begin{tabular}{l}
{\footnotesize * 2 sigma of the noise level.}\\
\end{tabular}
\end{center}
\end{table}

\clearpage

\begin{figure}
\begin{center}
\FigureFile(150mm,){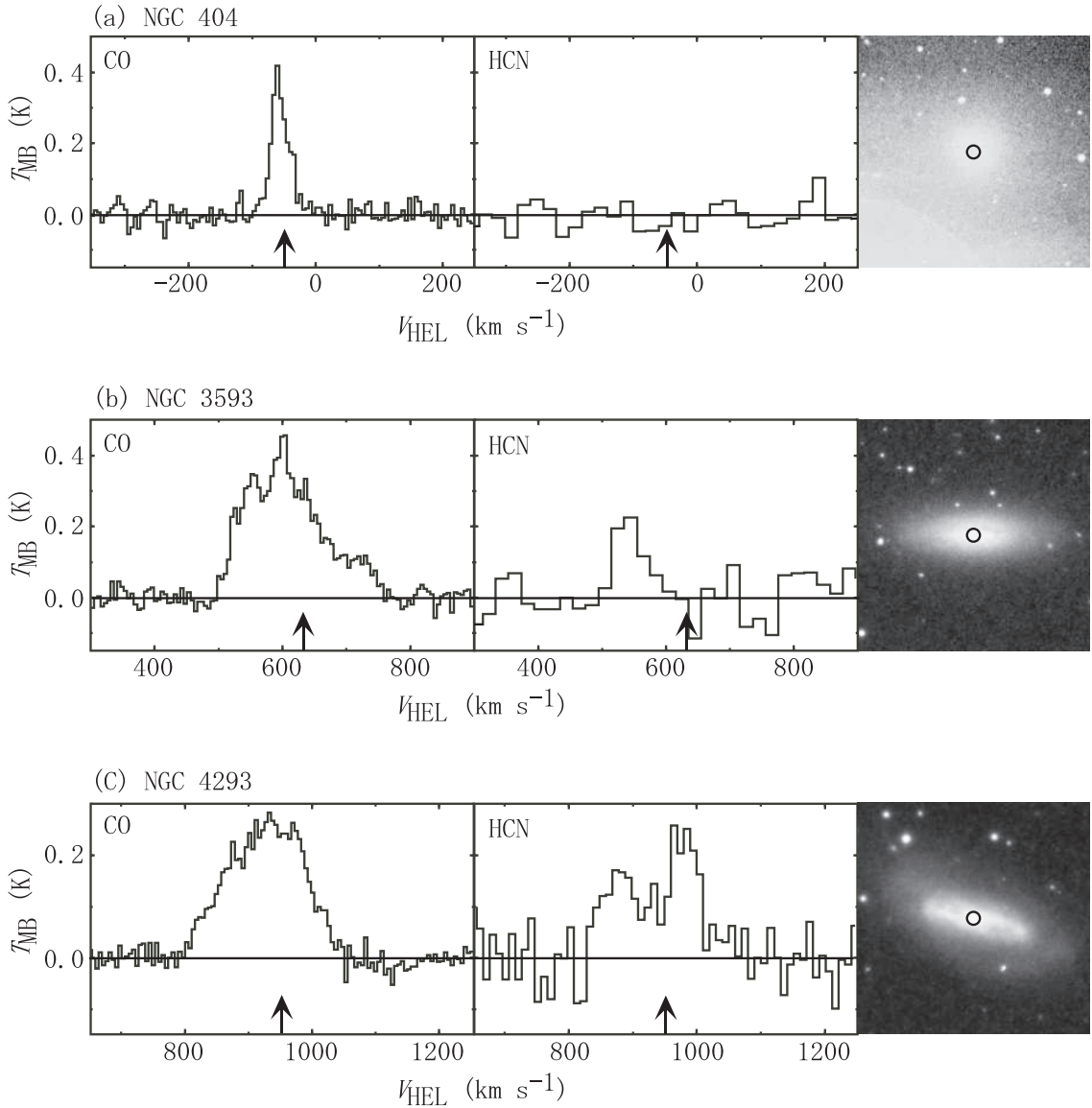}
\end{center}
\caption{Spectra of CO and HCN emission of (a) NGC 404, (b) NGC 3593, and (c) NGC 4293. The vertical scales of the HCN spectra have been multiplied by a factor of 10. All CO spectra have been smoothed to 5 km s$^{-1}$ resolution. HCN spectra have been smoothed to 20 km s$^{-1}$ resolution for NGC 404 and NGC 3593, and 10 km s$^{-1}$ resolution for NGC 4293. The arrows indicates the systemic velocity of the galaxies. Optical images are taken from the Digitized Sky Survey$^{1}$. The size of the images is $340'' \times 340''$. Circles are the beam size for HCN observations (19$''$). The top is the north.}
{\footnotesize $^{1}$The Digitized Sky Survey was produced at the Space 
Telescope Science Institute under U.S. Government grant NAG 
W-2166. The images of these surveys are based on photographic 
data obtained using the Oschin Schmidt Telescope on Palomar 
Mountain and the UK Schmidt Telescope. The plates were 
processed into the present compressed digital form with the 
permission of these institutions.}
\end{figure}

\clearpage

\begin{figure}
\begin{center}
\FigureFile(150mm,){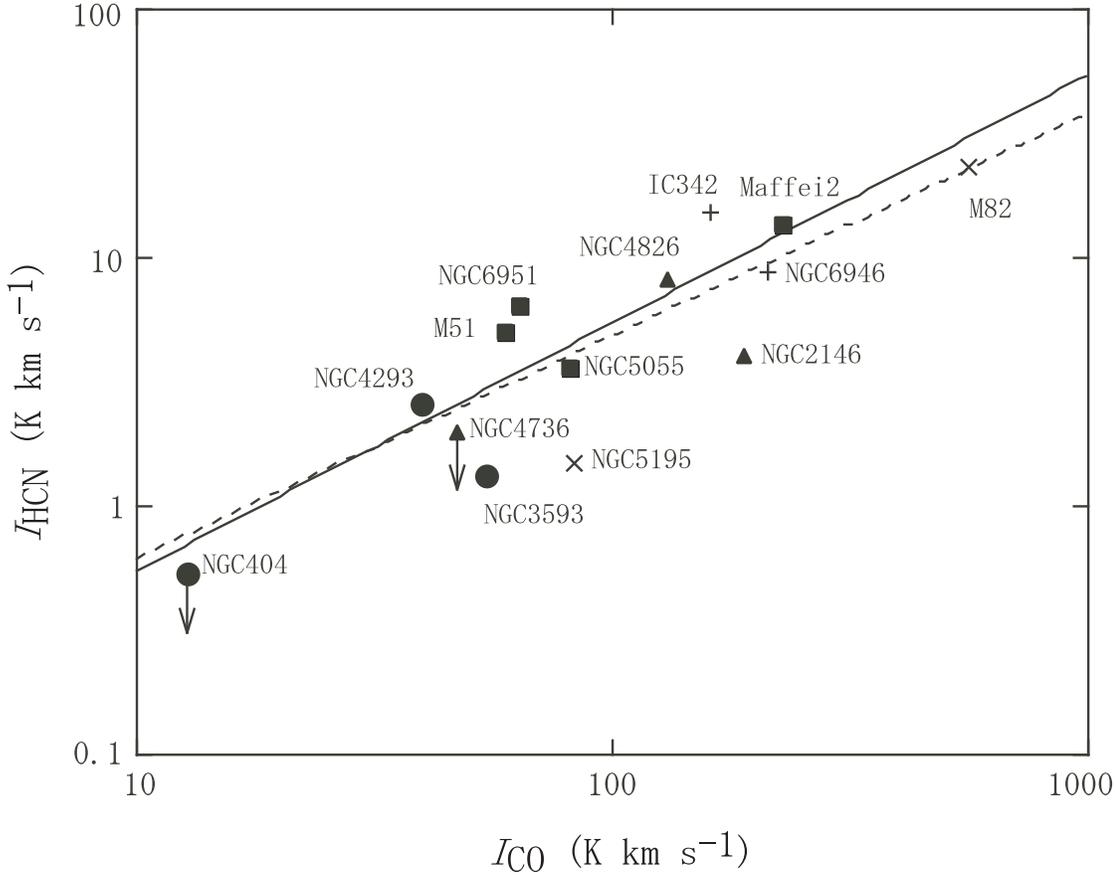}
\end{center}
\caption{Comparison between $I_{\rm CO}$ and $I_{\rm HCN}$. 
Different symbol indicates different Hubble type in Tully (1988): Filled circle indicates S0-S0/a, filled triangle Sab, filled square Sbc, plus sign Scd, cross P. 
The solid line indicates $I_{\rm CO}$/$I_{\rm HCN}$ = 0.055. 
The dashed line indicates $I_{\rm HCN} \propto I_{\rm CO}^{0.9}$.
The data of M82, IC342, Maffei2, NGC 2146, NGC6946 and M51 are from Sorai et al. (2002). 
The data of NGC 4736, NGC 4826, NGC 5055, and NGC 6951 are from Kohno (1998).
The data of NGC 5195 is from Kohno et al. (2002).}
\end{figure}

\end{document}